\documentclass[twocolumn,twoside,slac_two]{revtex4}
\usepackage{graphicx}
\usepackage{fancyhdr}

\usepackage{longtable}
\usepackage{amsfonts}
\usepackage{amsmath}
\usepackage{amssymb}
\usepackage{xspace}

\usepackage[colorlinks=true, citecolor=black, linkcolor=black, urlcolor=black, pdfauthor={Giovanni Piano}]{hyperref}

\newcommand{\de}{\mathrm{d}}

\renewcommand{\rho}{\varrho}

\def \gray {$\gamma$-ray\xspace}
\def \grays {$\gamma$-rays\xspace}
\def \flx {photons $\mathrm{cm}^{-2}$ $\mathrm{s}^{-1}$\xspace}
\def \grid {AGILE-\textit{GRID}\xspace}

\def \lat {\textit{Fermi}-LAT\xspace}

\def \hp {1AGL J2021+3652\xspace}
\def \gcyg {1AGL J2022+4032\xspace}
\def \srcyg {1AGL J2032+4102\xspace}
\def \hpsr {PSR J2021+3651\xspace}
\def \gcygpsr {PSR J2021+4026\xspace}
\def \srcygpsr {PSR J2032+4127\xspace}
\def \cyg {\mbox{Cygnus X-3}\xspace}

\begin{document}

\title{Transient Gamma-ray Emission from Cygnus X-3 Detected by AGILE: Leptonic and Hadronic Emission Models}

%

\author{G.~Piano}
\affiliation{INAF/IAPS, via del Fosso del Cavaliere 100, I-00133 Roma, Italy}
\author{M.~Tavani}
\affiliation{INAF/IAPS, via del Fosso del Cavaliere 100, I-00133 Roma, Italy}
\author{V.~Vittorini}
\affiliation{INAF/IAPS, via del Fosso del Cavaliere 100, I-00133 Roma, Italy}
\author{A.~Giuliani}
\affiliation{INAF/IASF-Milano, via E. Bassini 15, I-20133 Milano, Italy}
\author{on behalf of the AGILE team}

\begin{abstract}
The AGILE satellite detected several episodes of transient \gray emission from Cygnus X-3. Cross-correlating the AGILE light curve with both X-ray and radio monitoring data, we found that the main events of \gray activity were detected while the system was in soft spectral X-ray states, that coincide with local and often sharp minima of the hard X-ray flux, a few days before intense radio outbursts. This repetitive temporal coincidence between the \gray transient emission and spectral state changes of the source turns out to be the spectral signature of high-energy activity from this microquasar. The \gray differential spectrum of Cygnus X-3 (100 MeV -- 3 GeV), which was obtained by averaging the data collected by AGILE during the \gray events, is consistent with a power law of photon index $\alpha=2.0~\pm~0.2$.
Finally, we examined leptonic and hadronic emission models for the \gray activity and found that both scenarios are valid. In particular, in the leptonic model -- based on inverse Compton scatterings of mildly relativistic electrons on soft photons from both the Wolf-Rayet companion star and the accretion disk -- the emitting particles may also contribute to the overall hard X-ray spectrum, possibly explaining the hard non-thermal power-law tail seen during special soft X-ray states in Cygnus X-3.
\end{abstract}

\maketitle

\thispagestyle{fancy}

\section{Introduction}

\cyg is a high-mass X-ray binary, whose companion star is a Wolf-Rayet (WR) star \citep{vankerk_92} with a strong helium stellar wind \citep{szo_zdzia_08}. The system is located at a distance of about 7-10 kpc \citep{ling_09}. The orbital period is 4.8 hours, as inferred from infrared \citep{becklin_73}, X-ray \citep{parsignault_72}, and \gray \citep{abdo_09} observations. Owing to its very tight orbit (orbital distance $d \approx 3 \times 10^{11}$ cm), the compact object is totally enshrouded in the wind of the companion star \citep{vilhu_09}. The nature of the compact object is still uncertain although a black hole scenario is favored \citep{szo_zdzia_08, szostek_08}.
In the radio band, the system shows strong flares (``\textit{major radio flares}'') reaching up to few tens of Jy. Radio observations at milliarcsec scales confirm emissions (at cm wavelengths) from both a core and a one-sided relativistic jet ($v \sim 0.81c$), with an inclination to the line-of-sight of $\lesssim14^{\circ}$ \citep{mioduszewski_01}. The radiation from the jet dominates the radio emission from the core during (and soon after) the  major flares \citep{tudose_10}.
\cyg exhibits a clear, repetitive pattern of (anti)correlations between radio and X-ray emission, and an overall anticorrelation between soft and hard X-ray fluxes \citep{mccollough_99,szostek_08}.\\
Firm detections of high-energy \grays (HE \grays: $>$100 MeV) from \cyg were published at the end of 2009: the AGILE (Astro-rivelatore Gamma a Immagini LEggero) team found evidence that strong \gray transient emission above 100 MeV coincided with special X-ray/radio spectral states \citep{tavani_09a}, and the \lat (Large Area Telescope) collaboration announced the detection of \gray orbital modulation \citep{abdo_09}.
The \gray emission is most likely associated with a relativistic jet \citep{tavani_09a, abdo_09, dubus_10, cerutti_11, zdziarski_12}, but the radiative process (leptonic or hadronic) is uncertain.\\
A possible leptonic scenario for \gray emission in \cyg was proposed by \citealp{dubus_10}: stellar ultraviolet (UV) photons from the WR star are Compton upscattered to HE by relativistic electrons accelerated in the jet.\\
A hadronic scenario accounting for \gray emission in microquasars was discussed by \citet{romero_03,romero_05}. Their model is based on the interaction of a mildly relativistic jet with the dense wind of the companion star, and the \gray emission is due to the decay of neutral pions ($\pi^{0}$) produced by $pp$ collisions.\\
Here we report an analysis of the whole monitoring of \cyg during the ``pointing'' mode data-taking of the AGILE satellite. We present a study of the \gray spectrum by assuming both leptonic and hadronic scenarios.\\
This study is extracted from the paper of \citet{piano_12}, where all the details about this work are presented.

\section{Observations}

The AGILE scientific instrument \citep{tavani_09b} is very compact and characterized by two co-aligned imaging detectors operating in the energy ranges 30 MeV--30 GeV (Gamma-Ray Imaging Detector, \textit{GRID}: \citealp{barbiellini_02, prest_03}) and 18--60 keV (Super-AGILE: \citealp{feroci_07}), as well as by both an anticoincidence system \citep{perotti_06} and a calorimeter \citep{labanti_06}.\\
Until mid-October 2009 AGILE had operated in ``pointing'' mode with fixed attitude; in November 2009, AGILE entered ``scanning mode'', which is characterized by a controlled rotation of the pointing axis.\\
During the ``pointing'' mode data-taking ($\sim$2.5 years), \grid was characterized by enhanced performances in the monitoring capability of a given source, especially in the energy band 100--400 MeV (see \citealp{bulgarelli_12a} for details). Owing to the different pointing strategies of the AGILE and Fermi satellites, the high on-source cumulative exposure (between 100 and 400 MeV) of the \grid may be fundamental in the observation of this particular source.

We report an analysis based on the \grid data collected between 2007 November 2 and 2009 July 29. During this period, AGILE repeatedly pointed at the Cygnus region for a total of $\sim$275 days, corresponding to a net exposure time of $\sim$11 Ms. 
We performed an analysis of the whole \grid data using a detection algorithm developed by the AGILE team to automatically search for transient \gray emission. The algorithm initially analyzed 140 maps, each related to a 2-day integration (non-overlapping consecutive time intervals). The time bins containing the peak \gray emission, with detection significances greater than 3$\sigma$, were identified. The analysis was subsequently manually refined to optimize the determination of the time interval of the \gray emission. The whole analysis was carried out with the \verb+Build 19+ version of the AGILE team software, using the \verb+FM3.119_2+ calibrated filter applied to the consolidated dataset with off-axis angles smaller than $40^{\circ}$.
We used a multi-source maximum-likelihood analysis (MSLA) to take into account the emission of the nearby \gray pulsars\footnote{The main characteristics of the persistent \gray sources that we used in the MSLA are reported in Table 1 of \citet{chen_piano_11}.} \hp (\hpsr), \gcyg (\gcygpsr), and \srcyg (\srcygpsr). In particular, the MSLA is fundamental to avoid contamination by the pulsar \srcygpsr, located at a distance of $\sim 0.5^{\circ}$. In this paper, we did not consider the off-pulse data for the nearby pulsar. Nevertheless, the MSLA accounted for the steady \gray emission from this source when calculating the significance and the flux of each \gray detection of \cyg. Moreover, we can exclude any substantial spectral contamination from the pulsar because the steady \gray emission from the pulsar\footnote{The steady \gray emission from the pulsar \srcygpsr, as detected by the \grid, is $F_{\gamma}^{PSR} = [37 \pm 4(stat) \pm 10\% (syst)] \times 10^{-8}$ \flx for photon energies above 100 MeV, see \citet{chen_piano_11} for details.} is much fainter than the mean flux of the active \gray emission from \cyg.\\
The main events of \gray activity, detected with a significance above $3\sigma$ ($\sqrt{TS} \geqslant 3$), are shown in Table \ref{cyg_x3_all_flares}.

\begin{table}[htb]
\begin{center}
\caption{Main events of \gray emission detected by the \grid in the period November 2007 - July 2009. All detections have a significance above $3\sigma$ ($\sqrt{TS} \geqslant 3$). \textit{Column 1}: period of detection in MJD; \textit{Column 2}: significance of detection; \textit{Column 3}: photon flux (above 100 MeV).} \label{cyg_x3_all_flares}

{\small
\begin{tabular}{c|c|c}

\hline\hline
MJD        &  $\sqrt{\mathrm{TS}}$  &  Flux [$10^{-8}$ photons $\mathrm{cm^{-2}}$ $\mathrm{s^{-1}}$] \\
\hline
54507.76 - 54508.46 &        3.7         &           264  $\pm$  104   \\
\hline                                                                                 
54572.58 - 54573.58 &        4.5         &           265  $\pm$   80   \\
\hline
54772.54 - 54773.79 &        3.1         &           135  $\pm$   56   \\
\hline
54811.83 - 54812.96 &        4.0         &           190  $\pm$   65   \\
\hline
55002.88 - 55003.87 &        3.8         &           193  $\pm$   67   \\
\hline
55025.05 - 55026.04 &        3.2         &           216  $\pm$   89   \\
\hline
55033.88 - 55035.88 &        3.6         &           158  $\pm$   59   \\
\hline\hline

\end{tabular}
}
\end{center}
\end{table}

\begin{figure}[!h]
 \begin{center}
    \includegraphics[width=7.5cm]{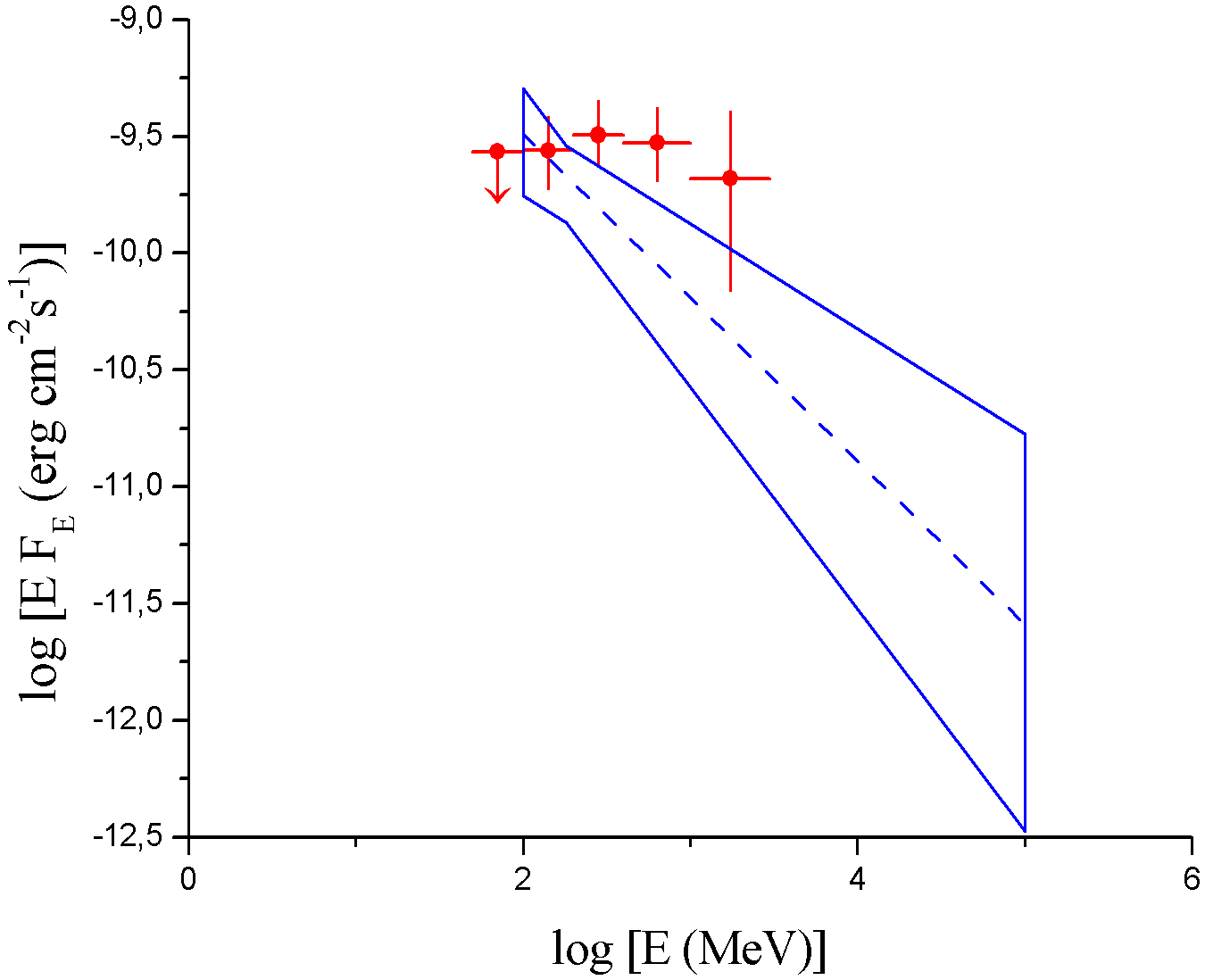}
    \caption{The $\nu F_{\nu}$ spectra of \cyg during the \gray activity. \textit{Red circles}: \grid energy spectrum (50 MeV to 3 GeV) of the main episodes. \textit{Blue error contours} and \textit{dashed blue line}: average power-law fit with $\alpha=2.70 \pm 0.25$ of the spectrum obtained by \lat integrating the two active windows of about two months each \citep{abdo_09}.} \label{cyg_x3_agile_fermi_only}
\end{center}
\end{figure}

We found seven events, including those presented in \citet{tavani_09a} and \citet{bulgarelli_12a}.
By integrating all the main events with the \verb+FM3.119_2+ filter, we detected a \gray source at $6.7\sigma$ ($\sqrt{TS}=6.7$) at the average Galactic coordinate $(l,~b) = (79.7^{\circ},~0.9^{\circ})$ $\pm~0.4^{\circ}$ (stat) $\pm~0.1^{\circ}$ (syst), with a photon flux of $(158 \pm 29) \times 10^{-8}$ \flx above 100 MeV\footnote{Here we present an updated result for the analysis of the 7-event integration with respect to the one reported in \citet{piano_11}. In this paper, our analysis was carried out with a more recent version of the AGILE software tool (AG\_multi4).}. The corresponding post-trial significance for repeated flare occurrences \citep{bulgarelli_12b} is 5.5 Gaussian standard deviations. The average differential spectrum between 100 MeV and 3 GeV is well-fitted by a power law with a photon index $\alpha=2.0~\pm~0.2$. 
In Figure~\ref{cyg_x3_agile_fermi_only}, we compare the $\nu F_{\nu}$ spectra of \cyg obtained by the \grid and \lat \citep{abdo_09} during the \gray activity. We remark that the \grid spectrum is related only to the peak \gray activity (the seven main events, lasting 1-2 days, in Table \ref{cyg_x3_all_flares}), whereas the \lat spectrum is an average spectrum found during the two active windows (of about two months each) of \gray emission from \cyg (MJD: 54750--54820 and MJD: 54990--55045).

In \citet{piano_12} we presented the comprehensive multiwavelength light curve of \cyg, showing the transient \gray activity detected by the \grid along with the hard X-ray (15-50 keV) fluxes from \textit{Swift}/BAT (Burst Alert Telescope), soft X-ray (3-5 keV) fluxes from \textit{RXTE}/ASM  (Rossi X-ray Timing Explorer/All Sky Monitor), and radio flux density from the AMI-LA (15 GHz) and RATAN-600 (2.15, 4.8, 11.2 GHz) radio telescopes. If we refer to the seven main \gray episodes reported in Table~\ref{cyg_x3_all_flares}, we found that \citep{piano_12}:

\begin{itemize}
  \item There is a \textit{strong anticorrelation} between the hard X-ray and \gray emission. The \gray flares detected by the \grid are coincident with minima of the hard X-ray light curve (\textit{Swift}/BAT count rate $\lesssim 0.02~\mathrm{counts}$ $\mathrm{cm^{-2}~s^{-1}}$).
  \item Every time we detect \gray activity, \cyg is in a soft spectral state (\textit{RXTE}/ASM count rate $\gtrsim 3~\mathrm{counts~s^{-1}}$).
  \item Every time we detect \gray episodes, the system is moving towards either a major radio flare (radio flux density $\gtrsim1$ Jy) or a quenched state preceding a major radio flare.
\end{itemize}

In particular, we found that the transient \gray emission occurs when the system is either moving into a quenched state (``pre-quenched'') or towards a radio flare (``pre-flare''), which has always been observed after a quenched state \citep{szostek_08}, i.e., the \gray emission is detected when the system is moving into or out of a quenched state. Hence, from a purely phenomenological point of view, the \textit{quenched state} seems to be a ``key'' condition  for the \gray emission.

Our study confirmed that the \gray emission conditions of \cyg, during the whole ``pointing'' monitoring by the AGILE satellite, are completely consistent with the ones found by \citealp{tavani_09a}, \citealp{bulgarelli_12a}, and \citealp{corbel_12}.

\section{Modeling the spectral energy distribution}

By accounting for the X-ray, \gray (\grid), and TeV emission (MAGIC spectral upper limits), we modeled the multiwavelength spectral energy distribution (SED) of \cyg during a soft spectral state, with both the \textit{leptonic} and \textit{hadronic} scenarios. We considered an X-ray spectrum measured by \textit{RXTE}-PCA\footnote{Proportional Counter Array (PCA)} and \textit{RXTE}-HEXTE\footnote{High Energy X-ray Timing Experiment (HEXTE)} ($\sim$3--150 keV) when the source was in a ``hypersoft'' state \citep{koljonen_10}, the \grid spectrum  for the main \gray events (red spectral points in Figure~\ref{cyg_x3_agile_fermi_only}), and the MAGIC differential flux upper limits obtained when the source was in the soft state. \citep{aleksic_10}. The hypersoft spectrum of \cyg is usually exhibited by the microquasar during the \textit{quenched state}, which has been found to be correlated to the transient \gray emission. This X-ray spectral state is characterized by a weak and hard power-law tail ($\alpha$= 1.7--1.9) of non-thermal origin.

\subsection{A leptonic scenario}

We modeled the multi-frequency SED by assuming a simple leptonic scenario in which a plasmoid of high energy electrons/positrons, injected into the jet structure, upscatters via inverse Compton interactions soft seed photons from both the WR star and the accretion disk. 

The physical parameters of the photon field are literature-based. We modeled the X-ray data with a black body (BB) spectrum characterized by a temperature $T_{bb}\approx1.3$ keV, which is consistent with the typical characteristic temperature of the disk during the hypersoft/ultrasoft state \citep{hjal_09,koljonen_10}, and a $L_{bb}\approx 8 \times 10^{37}~\mathrm{erg~s^{-1}}$. The main parameters that we used for the WR star are $T_{\star}=10^5~\mathrm{K}$ and $L_{\star} \approx 10^{39}~\mathrm{erg~s^{-1}}$ (see \citealp{dubus_10}). The WR star is assumed to emit UV photons isotropically. We modeled the average \gray emission in the orbital phase. Thus, the WR photons are assumed to come mainly from the side of the jet and collide with the relativistic leptons via IC scattering processes.

We carried out two different models: in the first one (leptonic model \textit{``A''}), the plasmoid interacts with the soft photon bath ``close'' to the disk (the star-plasmoid distance is $R \approx d \approx 3 \times 10^{11}$ cm), whereas in the second one (leptonic model \textit{``B''}) the interaction region is ``far away'' from the accretion disk (the star-plasmoid distance is $R \approx 10d \approx 3 \times 10^{12}$ cm).\\
For both models, the inclination of the jet to the line of sight is assumed to be $i = 14^{\circ}$, and the plasmoid is assumed to be spherical (radius $r = 3 \times 10^{10}$ cm) with a bulk motion characterized by a Lorentz factor of $\Gamma=1.5$ ($v = \sqrt{5}
c/3$). The population of electrons is modeled by a broken-power-law spectral distribution, with spectral indices $\alpha_1=2.2$, $\alpha_2=4.0$, $\gamma_{min}=1$, $\gamma_{max}=10^5$, and an energy break of $\gamma_b=4\times 10^3$. So that, for $\gamma_{min} \leqslant \gamma \leqslant \gamma_{max}$, we assumed that
\begin{equation} \label{broken_plaw}
 \frac{\de N}{\de \gamma \de \mathrm{V}}=\frac{K_e \, \gamma_b^{-1}}{\left(\frac{\gamma}{\gamma_b}\right)^{\alpha_1} + \left(\frac{\gamma}{\gamma_b}\right)^{\alpha_2}}.
 \end{equation}
The spectral indices and the energy break of the electron distribution are the best-fit values for the \grid spectral shape. The distribution of electrons/positrons is assumed to be isotropic in the plasmoid rest frame (the jet comoving frame).
We adopted the Klein-Nishina formula to describe the Compton scattering of soft photons by a cloud of mildly relativistic leptons \citep{aharonian_81}.

\begin{figure}[!tbp]
 \begin{center}
    \includegraphics[width=8.5cm]{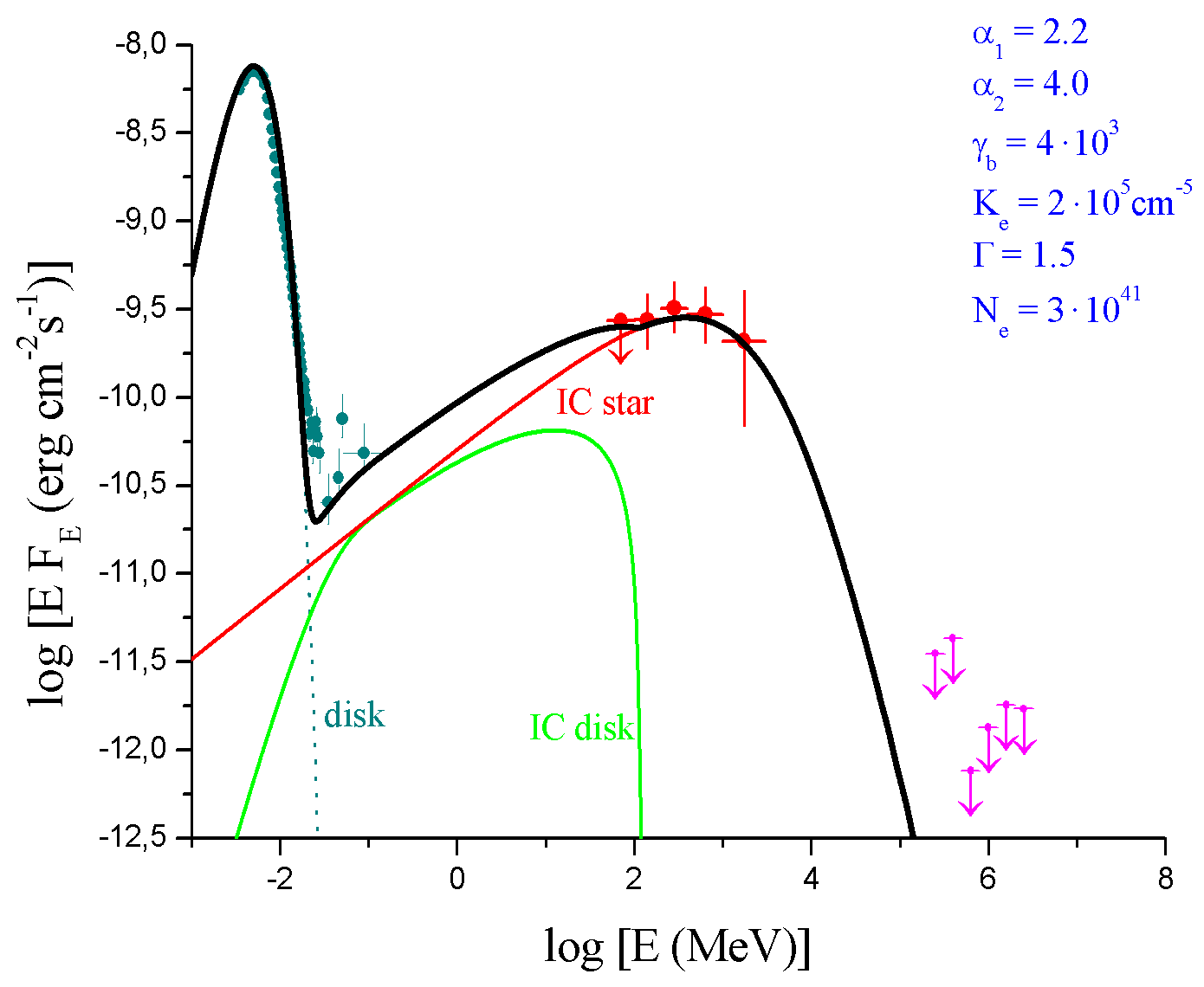}
    \caption{Multiwavelength SED of \cyg during the main \gray events (non-simultaneous data) and the leptonic model \textit{``A''} (see main text). \textit{Blue circles}: X-ray average ``hypersoft'' spectrum \citep{koljonen_10}, \textit{RXTE}-PCA and \textit{RXTE}-HEXTE data ($\sim$3 to $\sim$150 keV); \textit{red circles}: \grid energy spectrum (50 MeV to 3 GeV) of the main \gray episodes; \textit{magenta arrows}: MAGIC differential flux upper limits (95\% C.L.), 199--3155 GeV, related to soft spectral state \citep{aleksic_10}. Spectral components of the model are the BB emission from the disk (blue short-dashed line), IC scattering of the soft photons from the accretion disk (green solid line), and IC scattering of the soft stellar photons (red solid line). The global SED model curve is indicated by a black solid line.} \label{cyg_x3_mw_spectrum_lep_a}
\end{center}
\end{figure}

In the leptonic model \textit{``A''}, the distance from the star to the plasmoid location is assumed to be $R \approx 3 \times 10^{11}$ cm ($R \approx d$), i.e., the plasmoid in the jet is very close both to the compact object and the accretion disk. The distance between the plasmoid center and the compact object is $H \approx 3 \times 10^{10}$ cm, i.e., $H \approx r$. The results of this modeling are presented in Figure~\ref{cyg_x3_mw_spectrum_lep_a}.\\
We took into account the $\gamma \gamma$ absorption (for $e^{\pm}$ pair production) of the IC \gray photons by the X-ray photons from the accretion disk. We assumed that the distribution of the disk photons is fully isotropized by the stellar wind in the observer frame. This implies that the \gray photosphere (i.e., where $\tau_{\gamma \gamma}\geqslant1$) has a radius of $\sim10^{10}$ cm \citep{cerutti_11}. With these assumptions, the lowest part of the plasmoid is within the \gray photosphere\footnote{Assuming that the photosphere radius is equal to the plasmoid radius, i.e., $3 \times 10^{10}$ cm, the fraction of the plasmoid volume inside the \gray photosphere is $\sim$32\%.}. The spectral component related to the IC scatterings of the disk photons (green curve) is actually produced in this region, very close to the disk, where the X-ray photon density as well as the optical depth is high. Since $\tau_{\gamma \gamma} > 1$, this component displays a sharp cut-off energy at $\sim$100 MeV (i.e., the threshold for $e^{\pm}$ production, given the characteristic energies of the disk photons). On the other hand, the spectral component related to the IC scatterings of the stellar wind photons (red curve) does not show any cut-off energy, because it is mainly produced in the farthest part of the plasmoid (outside the \gray photosphere), where the $\gamma \gamma$ absorption by the X-ray disk photons is negligible. Thus, we deduced that in our geometry the plasmoid volume outside the \gray photosphere (distances $\gtrsim 10^{10}$ cm from the compact object) emits the bulk of the \gray emission above 100 MeV via IC processes acting on stellar photons. Conversely, we found that the innermost part of the jet (distances $\lesssim 10^{10}$ cm from the compact object), could contribute significantly to the hard X-rays at $\sim$100 keV (see Figure~\ref{cyg_x3_mw_spectrum_lep_a}).\\
In model \textit{``A''}, we found that the resulting lepton injection rate is $\dot{N}_e \approx 2 \times 10^{41}~\mathrm{leptons~s^{-1}}$ and the corresponding jet kinetic luminosity for the leptons ($L_{kin,~e} = \dot{N}_e~\Gamma~m_{e} c^2$) is $L_{kin,~e}^{A} \approx 2 \times 10^{35}$ $\mathrm{erg~s^{-1}}$.

\begin{figure}[!tbp]
 \begin{center}
    \includegraphics[width=8.5cm]{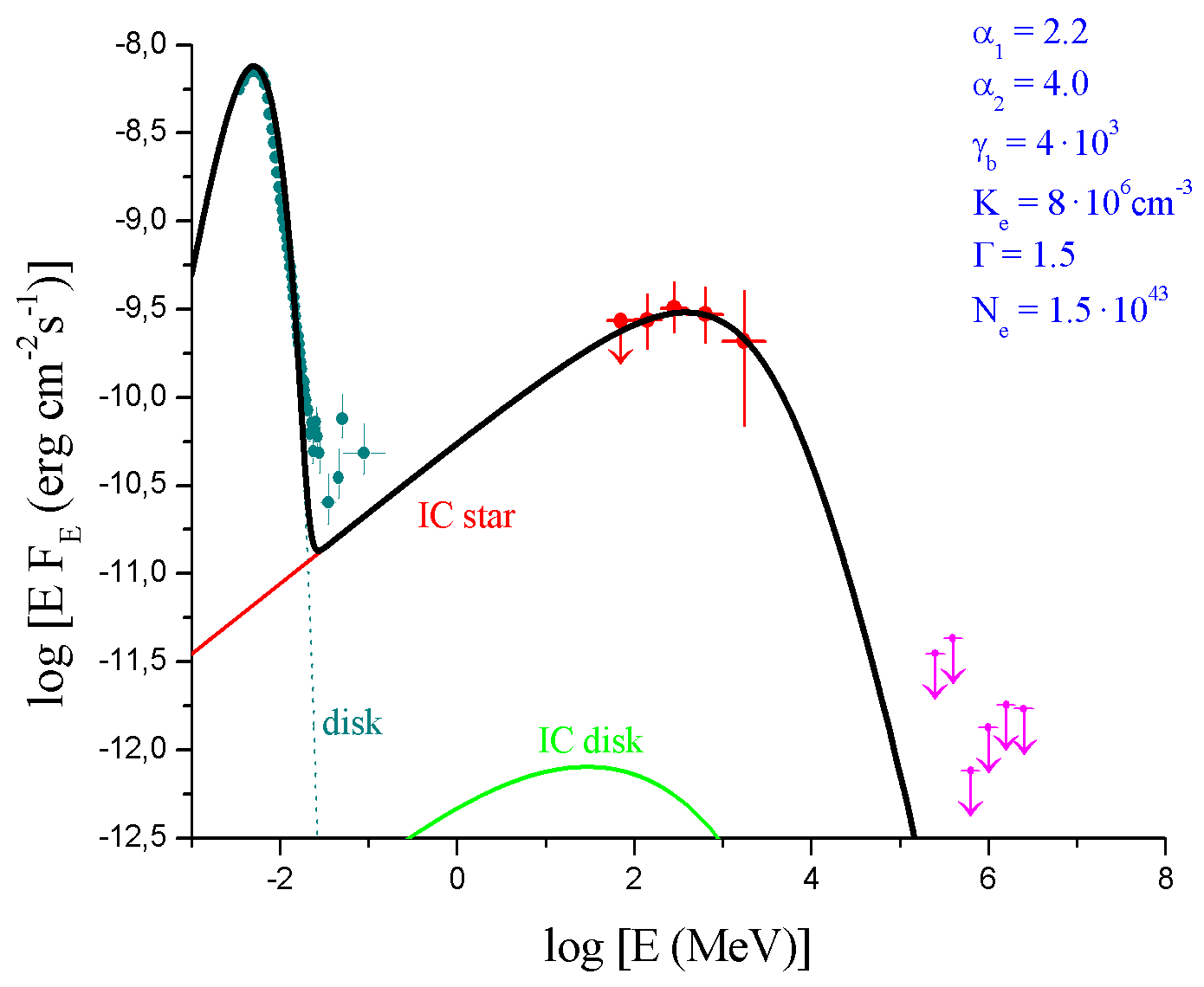}
    \caption{Multiwavelength SED of \cyg during the main \gray events (non-simultaneous data) and the leptonic model \textit{``B''} (see main text). Spectral components of the model are the BB emission from the disk (blue short-dashed line), IC scattering of the soft photons from the accretion disk (green solid line), and IC scattering of the soft stellar photons (red solid line). The global SED model curve is indicated by a black solid line. For a detailed description of the datasets, see caption to Figure~\ref{cyg_x3_mw_spectrum_lep_a}.} \label{cyg_x3_mw_spectrum_lep_b}
\end{center}
\end{figure}

In the leptonic model \textit{``B''}, the distance from the star to the plasmoid is assumed to be $R \approx 3 \times 10^{12}$ cm ($R \approx 10d$), i.e., the plasmoid in the jet is far away from the compact object and the accretion disk. The distance between the plasmoid center and the compact object is $H \approx 3 \times 10^{12}$ cm, i.e., $R \approx H$. We assumed that the disk photons enter the plasmoid mainly from behind. The results of this modeling are shown in Figure~\ref{cyg_x3_mw_spectrum_lep_b}. In this model, the spectral component related to the IC scatterings of disk photons (green curve) is negligible compared to the IC component of soft photons from the star (red curve).
We note that the ``IC disk'' component does not show any cut-off energy related to the $\gamma \gamma$ absorption by X-ray photons, because the IC \grays are produced well outside the \gray photosphere (at distances $\gg 10^{10}$ cm).\\
In model \textit{``B''}, we found that the resulting lepton injection rate is $\dot{N}_e \approx 10^{43}~\mathrm{leptons~s^{-1}}$ and the jet corresponding kinetic luminosity for the leptons is $L_{kin,~e}^{B} \approx 10^{37}$ $\mathrm{erg~s^{-1}}$.

\subsection{A hadronic scenario}

We also considered a ``hadronic scenario'' for \gray production
from \cyg based on the same formalism adopted by \citet{romero_03}.
In this case, the compact source is assumed to
eject a flux of mildly relativistic hadrons (mostly protons) at
the base of the jet. These protons are first accelerated near the
compact object and then propagate along the jet interacting with
the gaseous surroundings provided by the WR companion mass-outflow.
The resulting proton-proton ($pp$) collisions can copiously
produce pions and \grays resulting from neutral pion decays.

The proton distribution in the jet is assumed to be isotropic in the jet comoving frame,
with an energy spectrum described by a power law with a high energy cut-off (for $\gamma \geqslant \gamma_{min}$)
\begin{equation} \label{plaw_coff}
 \frac{\de N}{\de \gamma \de \mathrm{V}}=K_p \, \gamma^{-\alpha}~{\rm exp}(-\gamma/\gamma_c)
\end{equation}
with $\alpha=3$, $\gamma_{min}=1$, and $\gamma_c=100$.
The spectral index of the distribution is the best-fit value for the \grid spectral shape. 
We set the energy cut-off value at $\gamma_c=100$ so that the total SED is consistent with the spectral constraints of the MAGIC upper limits.

The ejected protons interact with the hadronic
matter of the WR strong wind. The inelastic hadronic scatterings produce neutral
pions that subsequently decay into \grays.
We adopted the same formula for the cross section $\sigma_{pp}(\gamma)$ of inelastic $pp$ interaction
reported by \citet{kelner_06}.
We assumed that the injected protons in the jet interact with the gas of the wind along a
cylindrical column of matter with a
radius $r = 3 \times 10^{10}$ cm and a height of $H \approx 3 \times
10^{12}$ cm. 
In analogy with the leptonic models, we assumed for the jet a bulk Lorentz factor of $\Gamma=1.5$, 
and an inclination to the line of sight of $i = 14^{\circ}$.
To quantify the density of matter in the WR wind, we assumed that
the companion star has a mass-loss rate of $\dot{M} \sim 10^{-5}
M_{\odot}~\mathrm{yr}^{-1}$ and the speed of the wind is $v_{wind}
\sim 1000~\mathrm{km~s^{-1}}$ \citep{szo_zdzia_08}. By integrating the
density of matter in this cylinder expressed in terms of the number
density of protons ($\rho \sim 1/R^2$, where $R$ is the distance
from the star), we find that the total number of protons from the
wind in this column is $N_{p,wind} \approx 3.7 \times 10^{45}$.

\begin{figure}[!tbp]
 \begin{center}
    \includegraphics[width=8.5cm]{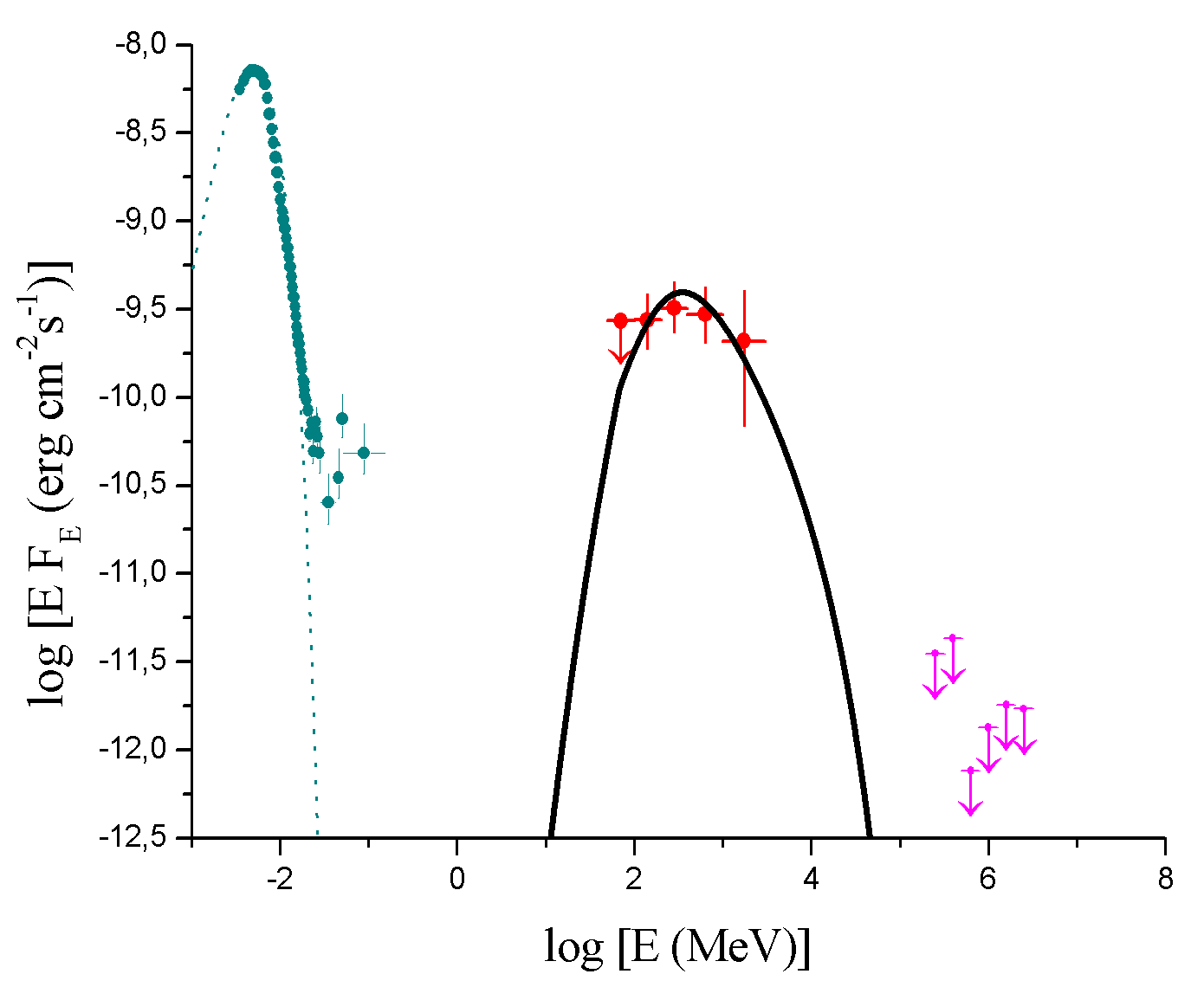}
    \caption{Multiwavelength SED of \cyg during \gray events (non-simultaneous data) and a hadronic model (see main text). Black body emission from the disk (blue short-dashed line), and \gray emission from $\pi^0$-decays (black solid line) are plotted. For a detailed description of the datasets, see caption to Figure~\ref{cyg_x3_mw_spectrum_lep_a}.} \label{cyg_x3_mw_spectrum_had}
\end{center}
\end{figure}

The result of our best-fit hadronic model for \cyg is reported
in Figure~\ref{cyg_x3_mw_spectrum_had}. In this case, we found a resulting proton injection rate of
$\dot{N}_{p,jet} \approx 6.7 \times 10^{40}$ protons $\mathrm{s^{-1}}$ and the corresponding jet kinetic luminosity for the hadrons ($L_{kin,~p} = \dot{N}_p~\Gamma~m_{p} c^2$) is  $L_{kin,~p} \approx 1.5 \times 10^{38}$ $\mathrm{erg~s^{-1}}$. This value is consistent with the average bolometric luminosity of the hypersoft state, $L^{HYS}_{bol} \approx 1.2 \times 10^{38}$ $\mathrm{erg~s^{-1}}$ \citep{koljonen_10} and it is lower than the Eddington accretion limit for the system ($L_{Edd} \approx 10^{39}$ $\mathrm{erg~s^{-1}}$, assuming that the compact object is a black hole with a mass of $M_x \approx 10 M_{\odot}$).

\section{Conclusions}

We found that the \gray spectrum of \cyg detected by the \grid is significantly harder than the time-averaged spectrum obtained by \lat for the ``\gray active periods'' of the microquasar, lasting $\sim$4 months (see Figure \ref{cyg_x3_agile_fermi_only}). Although both the AGILE main \gray events and the Fermi \gray active periods are both likely related to the presence of an active jet, the spectral difference may imply that there was a fast hardening of the spectrum during the peak \gray events, lasting $\sim$1-2 days.

We have demonstrated that both a leptonic model -- based on inverse Compton emission from a relativistic plasmoid injected into the jet -- and a hadronic model -- based on $\pi^0$-decays -- might account for the \gray emission observed by the \grid. Both of these models require the introduction of a new component (``IC bump'' or ``$\pi^0$-bump'') into the SED of the system. In both the leptonic and hadronic pictures, the inclination of the jet to the line of sight is assumed to be $i = 14^{\circ}$.

A leptonic scenario seems to be more likely than a hadronic one: the \gray modulation, the spectral link between hard X-ray and \gray spectra, and the temporal link between \gray events and radio flares could be interpreted in a natural way by assuming that the electrons are the main emitters. According to our results, the innermost part of the jet (distances $\lesssim 10^{10}$ cm from the compact object) could provide a strong contribution to the hard X-rays at $\sim$100 keV during  the \gray emitting interval, while the farthest part (distances between $10^{10}$ cm and $10^{12}$ cm from the compact object) produces the bulk of the \gray emission above 100 MeV.

In the context of a hadronic scenario, we found that a simply hadronic model can account for the \gray spectrum detected by the \grid, by assuming a reasonable proton injection rate in the jet. Our model, with the assumption of a standard WR wind, would require a jet kinetic power that is lower than the Eddington accretion limit for a black hole with a mass of $M_x \approx 10 M_{\odot}$. Thus, a hadronic picture is physically reasonable and not energetically less likely than a leptonic one. Hadronic mechanisms, besides emitting strong \gray radiation via $\pi^0$-decay, would produce an intense flux of high-energy neutrinos, emerging from the decay of secondary charged mesons produced in $pp$ collisions. Hence, a firm simultaneous detection of strong neutrino flux and \gray activity from \cyg would represent the signature of a dominant hadronic mechanism in the relativistic jet. At present, there is no strong evidence that one of these hypotheses can be excluded, and it remains an open question whether the dominant process for \gray emission in microquasars is either hadronic or leptonic \citep{mirabel_12}.


\begin{thebibliography}{99} 

{\small

\bibitem[Abdo et al.(2009)]{abdo_09} Abdo, A.~A., et al.\ 2009, Science, 326, 1512

\bibitem[Aharonian \& Atoyan(1981)]{aharonian_81} Aharonian, F.~A., \& Atoyan, A.~M.\ 1981, Ap\&SS, 79, 321 

\bibitem[Aleksi{\'c} et al.(2010)]{aleksic_10} Aleksi{\'c}, J., et al.\ 2010, ApJ, 721, 843

\bibitem[Barbiellini et al.(2002)]{barbiellini_02} Barbiellini, G., et al.\ 2002, NIMPA, 490, 146

\bibitem[Becklin et al.(1973)]{becklin_73} Becklin, E.~E., et al. \ 1973, Nature, 245, 302

\bibitem[Bulgarelli et al.(2012a)]{bulgarelli_12a} Bulgarelli, A., et al.\ 2012a, A\&A, 538, A63 

\bibitem[Bulgarelli et al.(2012b)]{bulgarelli_12b} Bulgarelli, A., et al.\ 2012b, A\&A, 540, A79 

\bibitem[Cerutti et al.(2011)]{cerutti_11} Cerutti, B., et al.\ 2011, A\&A, 529, A120 

\bibitem[Chen, Piano et al.(2011)]{chen_piano_11} Chen, A.~W., Piano, G., et al.\ 2011, A\&A, 525, A33

\bibitem[Corbel et al.(2012)]{corbel_12} Corbel, S., et al.\ 2012, MNRAS, 421, 2947

\bibitem[Dubus et al.(2010)]{dubus_10} Dubus, G., et al.\ 2010, MNRAS, 404, L55

\bibitem[Feroci et al.(2007)]{feroci_07} Feroci, M., et al.\ 2007, NIMPA, 581, 728

\bibitem[Hjalmarsdotter et al.(2009)]{hjal_09} Hjalmarsdotter, L., et al.\ 2009, MNRAS, 392, 251

\bibitem[Kelner et al.(2006)]{kelner_06} Kelner, S.~R., et al.\ 2006, Phys.~Rev.~D., 74, 034018 

\bibitem[Koljonen et al.(2010)]{koljonen_10} Koljonen, K.~I.~I., et al.\ 2010, MNRAS, 406, 307

\bibitem[Labanti et al.(2006)]{labanti_06} Labanti, C., et al.\ 2006, Proc.~SPIE, 6266

\bibitem[Ling et al.(2009)]{ling_09} Ling, Z., et al.\ 2009, ApJ, 695, 1111

\bibitem[McCollough et al.(1999)]{mccollough_99} McCollough, M.~L., et al. \ 1999, ApJ, 517, 951

\bibitem[Mioduszewski et al.(2001)]{mioduszewski_01} Mioduszewski, A.~J., et al.\ 2001, ApJ, 553, 766

\bibitem[Mirabel(2012)]{mirabel_12} Mirabel, I.~F.\ 2012, Science, 335, 175 

\bibitem[Parsignault et al.(1972)]{parsignault_72} Parsignault, D.~R., et al.\ 1972, Nature, 239, 123

\bibitem[Perotti et al.(2006)]{perotti_06} Perotti, F., et al.\ 2006, NIMPA, 556, 228

\bibitem[Piano et al.(2011)]{piano_11} Piano, G., et al.\ 2011, arXiv:1110.6043 

\bibitem[Piano et al.(2012)]{piano_12} Piano, G., et al.\ 2012, A\&A, 545, id.A110

\bibitem[Prest et al.(2003)]{prest_03} Prest, M., et al.\ 2003, NIMPA, 501, 280

\bibitem[Romero et al.(2003)]{romero_03} Romero, G.~E., et al.\ 2003, A\&A, 410, L1 

\bibitem[Romero et al.(2005)]{romero_05} Romero, G.~E., et al.\ 2005, ApJ, 632, 1093 

\bibitem[Szostek \& Zdziarski (2008)]{szo_zdzia_08} Szostek A., Zdziarski A. A., 2008, MNRAS, 386, 593S

\bibitem[Szostek et al.(2008)]{szostek_08} Szostek, A.,  et al.\ 2008, MNRAS, 388, 1001

\bibitem[Tavani et al.(2009a)]{tavani_09a} Tavani, M., et al.\ 2009a, Nature, 462, 620

\bibitem[Tavani et al.(2009b)]{tavani_09b} Tavani, M., et al.\ 2009b, A\&A, 502, 995

\bibitem[Tudose et al.(2010)]{tudose_10} Tudose, V., et al.\ 2010, MNRAS, 401, 890

\bibitem[van Kerkwijk et al.(1992)]{vankerk_92} van Kerkwijk, M.~H., et al.\ 1992, Nature, 355, 703

\bibitem[Vilhu et al.(2009)]{vilhu_09} Vilhu, O.; et al. \ 2009, A\&A, 501, 679

\bibitem[Zdziarski et al.(2012)]{zdziarski_12} Zdziarski, A.~A., et al.\ 2012, MNRAS, 421, 2956 

}
\end{thebibliography}

\end{document}